\documentclass[10pt,a4paper,onecolumn]{article}

\usepackage{microtype}
\usepackage{science}
\usepackage{amsmath}
\usepackage{amssymb}
\usepackage{wasysym}
\usepackage[italian,english]{babel}
\usepackage{amsthm}
\usepackage{amscd}
\pdfoutput=1 

\usepackage[colorlinks,hyperindex,plainpages=false]{hyperref}
\usepackage{hypcap}
\usepackage{thumbpdf}
\usepackage{url}

\theoremstyle{plain}
\newtheorem{thm}{Theorem}[section]

\theoremstyle{definition}
\newtheorem{defn}{Definition}[section]

\theoremstyle{remark}
\newtheorem*{rem}{Remark}

\usepackage{thumbpdf}

\begin{document}
\title{Seasonal Effects on Honey Bee Population Dynamics: a Nonautonomous System of Difference Equations}

\author{Gianluca Gabbriellini}
\maketitle
\thispagestyle{empty}

\begin{abstract}

The honey bees play a role of unquestioned relevance in nature and the comprehension of the mechanisms affecting their population dynamic is of fundamental importance. As experimentally documented, the proper development of a colony is related to the nest temperature, whose value is maintained around the optimal value if the colony population is sufficiently large. Then, the environmental temperature, the way in which this influence the nest temperature and the colony population size, are variables closely linked to each other and deserve to be taken into account in a model that aims to describe the population dynamics. In the present study, as first step, the continuous-time autonomous system proposed by Khoury, Myerscoug and Barron (KMB) in 2011 was approximated by means a Nonstandard finite difference (NSFD) scheme in order to obtain a set of autonomous difference equations. Subsequently, with the aim to introduce the seasonal effects, a nonautonomous version (NAKMB) was proposed and formulated in discrete-time domain via a NSFD scheme, by introducing a time–-dependent formulation for the queen bee laying rate and the recruitment rate coefficients. By means the phase-plane analysis was possible to deduce that, with an appropriate choice of the parameters, the NAKMB model admits both a limit cycle at nonzero population size and an equilibrium point marking the colony collapse, depending on the initial population size. 

\end{abstract}

\section{Introduction}

The honey bees are perhaps the most studied insects because their pollinating activities have a fundamental impact on the whole
ecosystem. The growing attention to the safeguard of honey bees requires to deep understand the mechanisms that impact on the life of a
colony. In order to rigorously describe the population dynamics of a colony and the role played by the surrounding environment, the mathematical
modeling could be important to address the problem. 

The honey bee colonies are composed by three castes: $20$-$40$ thousand workers, a queen and zero to few thousand drones \cite{page2001}. The
drone bee is a male and his main function is to be ready to fertilize a receptive queen, that is the unique responsible for laying eggs and
for this reason is the parent of the whole colony. The worker bees are infertile and their energies are completely dedicated to the survival
of the colony, by serving many roles during their lifetime: colony maintenance, brood rearing tasks, defense and foraging, among the main roles. A complete discussion about the roles of the worker bees can be found in \cite{johnson2010}. For the purposes of this paper, it is useful to classify the work of the bees in two categories, according to the level of risk to which they are subjected: \textit{hive bees} and \textit{foragers}. The firsts live in a protected environment, the seconds are exposed to the external climatic conditions and are constantly life-threatening. The model
proposed by Khoury, Myerscoug and Barron in 2011 \cite{khoury2011}
(cited in the present work as KMB model) describes a demographic model to explore the process of colony failure, by discussing the effect of different death rates of forager bees on colony growth. The model forecasts a threshold forager death rate $m^*$ such that if $m>m^*$ the colony is doomed to failure. The model proposed by Brown in 2013 \cite{brown2013} introduces a nonzero death rate also for hive bees, knowingly neglected in \cite{khoury2011}. In both models the authors study a system of \textit{autonomous} differential equations: none of the parameters explicitly depends on time.

Nevertheless, the seasonal effects are of a fundamental importance and deserve to be considered so that the model can
describe the population dynamics as realistically as possible. The seasons are marked by changes in weather, in particular the temperature is one of
the most impacting factors regulating the life cycle of all animals. In this paper were taken into account only the effects related to
the annual variations (seasonal), neglecting the circadian ones, also
important as discussed, i.e., in \cite{fuchikawa2007}. Seasonal
effects have been investigated by Russell, Barron and Harris in
2013 \cite{russell2013}, with a detailed dynamic flow model using a
commercial software (\textit{Stella}, isee system - version 8.0) that takes into account
the influence of the seasonality on death rates and food
availability. Their work report also a complete review of the other models taking into account seasonal effects.   

In this paper is first proposed a discrete-time version of the KMB model, by taking advantage of the \textit{Nonstandard Finite Difference Scheme} (NSFD) \cite{mickens89}. The NSFD introduces some rules ensuring a finite difference scheme without the instabilities sometimes introduced by the other discretization methods. In scientific literature there are many applications of the NSFD scheme, each showing the robustness and flexibility of the method (see, i.e., \cite{mickens89,mickens94,mickens00,mickens01,gabbriellini2014}).
After, the seasonal effects were evaluated by introducing a new formulation for the queen laying rate and the recruitment functions. The problem is first formulated in continuous-time domain by means a system of two nonlinear nonautonomous differential
equations and after, applying the NSFD rules, in discrete-time domain.
 
\section{Definitions and preliminaries} 

In this section some basic definitions will be given, about the concept of stability of a dynamical system, with particular regard to discrete-time formulation.

\subsection{Autonomous dynamical systems} \label{sub:AutDynSyst}

\subsubsection{Continuous-time}
A  general $n$-dimensional \textit{autonomous continuous-time dynamical system} is defined by the equation

\begin{align}\label{eq:differential_equation}
\left\{
\begin{array}{l}
\frac{d\xi(t)}{dt}=F(\xi(t),k) \\ \\
\xi(t_0)=\xi_0\in \mathbb{R}^n_+, \\
\end{array}
\right.
\end{align}
with $F: \mathbb{R}^n\longrightarrow \mathbb{R}^n$ is supposed to be differentiable in $\mathbb{R}^n_+$ and constitutes a vector field, $\xi=\xi(t):[0,+\infty)\longrightarrow \mathbb{R}^n$ is the state at time $t$, with initial condition $\xi_0\in\mathbb{R}^n_+$, and $k=(k_1,k_2,\ldots)$ represents the system parameters. If $F$ is globally Lipschitz, then there exists a unique solution $\xi(t)$ for all $t\ge t_0$, hence the \eqref{eq:differential_equation} defines a dynamical system on $\mathbb{R}^n$. In general $F$ could be linear or nonlinear respect to $\xi$; in this paper only nonlinear systems will be discussed. 

\begin{defn}
For a system defined with \eqref{eq:differential_equation}, a \textit{steady-state} solution is a point $\tilde\xi\in\mathbb{R}^n$ satisfying the relation:
\begin{align}\label{eq:fixed_point_cont}
F(\tilde\xi)=0.
\end{align}
\end{defn}
In the continuation of the text the set $\Gamma_c=\{\tilde\xi|f(\tilde\xi)=0,\,\tilde\xi\in\mathbb{R}^n\}$ will indicate the set of steady-states solutions.

In order to evaluate the stability of the steady-state solutions of the nonlinear system, a linearization around each equilibrium point is required, through the calculation of the Jacobian matrix, indicated with $J_F$, and its eigenvalues, indicated with $\lambda$.
\begin{defn}\label{def:stab_cont}
Let $\sigma(J_F)$ the set of eigenvalues of $J_F$, a steady-state $\tilde\xi\in\mathbb{R}^n$ for which $J_F$ has no eigenvalues with zero real parts, named \textit{hyperbolic steady-states}, is \cite{lyapunov1992}:
\begin{itemize}
\item \textit{asymptotically stable} if and only if $\Re\lambda<0$, for all $\lambda\in\sigma(J_F)$,
\item \textit{unstable} if and only if $\Re\lambda>0$, for all $\lambda\in\sigma(J_F)$.
\end{itemize}
This criterion does not apply for \textit{nonhyperbolic steady-states}, characterized by one or more eigenvalues with $\Re\lambda=0$.
\end{defn}
The correspondence between the behavior of the nonlinear equations and the linearized version is ensured by the Hartman--Grobman theorem.
\begin{thm}[Hartman--Grobman]\label{th:HG}
If $\tilde\xi$ is a hyperbolic equilibrium of $d\xi(t)/dt=F(\xi(t))$, then there is a neighborhood of $\tilde\xi$ in which $F$ is topologically equivalent\footnote{Two dynamical systems $d\xi(t)/dt=F_1(\xi(t))$ and $d\xi(t)/dt=F_2(\xi(t))$ defined on open sets $U$ and $V$ of $\mathbb{R}^2$ are topologically equivalent if there exists a homeomorphism $\alpha:U\longrightarrow V$ mapping the orbits of $F_1$ onto those of $F_2$ and preserving direction in time \cite{epstein1997}.} to the linear vector field $d\xi(t)/dt=J_F(\tilde\xi)\xi$.
\end{thm}

\subsubsection{Discrete-time}
In case of an autonomous discrete-time model, the continuous variable $t$ must be
replaced by $t_{0},t_1,\ldots,t_k$, with $t_k-t_{k-1}=\Delta t$, in which
$\Delta t$ is the constant time-step; the variable $\xi(t)$ must take
discrete values $\xi_n$. Then, the differential equation becomes a
difference equation. 

Let $f:\mathbb{R}^n\longrightarrow\mathbb{R}^n$,
consider a sequence $\{\xi_n\}_{n=0}^{\infty}$: it can be defined by a
mapping
$~\Lambda:\mathbb{R}^n\times\mathbb{R}^n\longrightarrow \mathbb{R}^n$
of the form $H(\xi_{n+1},\xi_{n})$. In some cases, is possible that
$\xi_{n+1}$ is given explicitly in terms of $\xi_{n}$:
\begin{align}\label{eq:diff_eq0}
\xi_{n+1}=\Phi(\xi_n,k),
\end{align}
where $\Phi:A\subseteq\mathbb{R}^n\longrightarrow\mathbb{R}^n$ and $k=(k_1,k_2,\ldots)$ represents the system parameters.

\begin{defn} 
A steady-state (or \textit{fixed point}) $\tilde\xi_n\in\mathbb{R}^n$ of \eqref{eq:diff_eq0}
respects the following conditions:
\begin{align}\label{eq:fixed_point_disc}
\Phi(\tilde\xi_n)=\tilde\xi_n.
\end{align}
\end{defn}
Likewise to the continuous case, also in discrete case it is useful to
indicate with $\Gamma_d=\{\tilde\xi_n|\Phi(\tilde\xi_n)=0,\,\tilde\xi_n\in\mathbb{R}^n\}$
the set of steady-states. 

\begin{defn}\label{def:stab_disc}
Let $\Phi\in C^1$, $J_{\Phi}$ its $n\times n$ Jacobian matrix and $\sigma(J_\Phi)$ the set of the Jacobian eigenvalues, a theorem (see i.e., \cite{mathmodecol}) ensures that a steady-state $\tilde\xi_n\in\mathbb{R}^n$ is:
\begin{itemize}
\item locally \textit{asymptotically stable} $\iff$ $\Re\lambda<1$, $\forall\,\lambda\in\sigma(J_\Phi)$: this point is an \textit{attractor};
\item \textit{unstable} $\iff$ $\Re\lambda>1$, $\forall\lambda\in\sigma(J_\Phi)$: this point is a \textit{repeller};
\item no conclusions on stability if $\Re\lambda>1$ for some $\lambda\in\sigma(J_\Phi)$. 
\end{itemize}
\end{defn}

\begin{defn}
The finite difference method is called \textit{elementary stable} if for all $\Delta t>0$, the stability properties of each $\tilde\xi_n\in\Gamma_d$ are the same of each $\tilde\xi\in\Gamma_c$.
\end{defn}

Let $\tilde J_\Phi:= J_\Phi(\tilde \xi_n)$, in two dimensional systems the characteristic polynomial of the Jacobian can be written as:
\begin{align}
\lambda^2-\lambda \operatorname{tr} (\tilde J_\Phi)+\det(\tilde J_\Phi).
\end{align}
In order to have $\Re\lambda<1$ for all $\lambda\in\sigma(J_\Phi)$, the Jury condition \cite{Jury} states that:
\begin{align}\label{eq:Jury}
|\operatorname{tr}(\tilde J_\Phi)|<1+\det(\tilde J_\Phi)<2.
\end{align}
Therefore, this criterion establishes that exists a necessary and sufficient condition to guarantee the asymptotic stability of the steady-state solutions. 

\subsection{Nonautonomous dynamical systems and the Poincar\'e map}\label{subsec:poincare}

\begin{defn}
If the vector field $F$ of \eqref{eq:differential_equation} depends explicitly on $t$, the system of equations is called \textit{nonautonomous}. 
\end{defn}

\begin{defn} In case of a nonautonomous dynamical system, if there exists a $\tau>0$ such that $F(\xi,t)=F(\xi,t+\tau)$ for all $\xi\in\mathbb{R}^n,\,t\in\mathbb{R}$, the system is said \textit{time-periodic} with period $\tau$. 
\end{defn}
A such $\tau$-periodic dynamical system can be converted into a $(n+1)$-order autonomous dynamical system by adding an equation \cite{parker1989}:  
\begin{align}\label{eq:differential_equation_na_1}
\left\{
\begin{array}{l}
\frac{d\xi(t)}{dt}=F(\xi(t),\theta\tau /2\pi,k) \\ \\
\frac{d\theta(t)}{dt}=\frac{2\pi}{\tau} \\ \\
\xi(t_0)=\xi_0\in \mathbb{R}^n_+,\,\theta(t_0)=2\pi t_0 /\tau. \\
\end{array}
\right.
\end{align}
In such a way the problem is formulated in the $\mathbb{R}^{n}\times S^1$ toroidal phase space, where $S:=[0,2\pi)$. If $\theta$ is substituted with $\theta+2\pi m$ for $m\in\mathbb{Z}$, then the system is unchanged. In this space, the planes identified by $\theta=2\pi m$ coincide with $\theta=0$, which is a section for a Poincar\'e sequence \cite{jordan2007}. 

\section{Rules to built a numerically stable finite difference scheme}\label{sec:rulesNSFD}

The numerical integration of ordinary differential equations using
traditional methods could produce different solutions from those of
the original ODE \cite{mickens05,gabbriellini2014,cresson16}. In
particular, using a discretization step-size larger than some relevant time
scale, is possible to obtain solutions that may not reflect the dynamics
of the original system \cite{mickens05}. To overcome this problem, Ronald
Mickens, in 1989, suggested what is known as the Nonstandard Finite
Difference (NSFD) method \cite{mickens89}, based on the concept of Dynamic Consistency. 

\begin{defn}
Let a first-order autonomous ODE and given $U$ its set of properties, the correspondent difference equation is \textit{dynamically consistent} with the ODE if it respects the same set of properties: stability, bifurcations and eventually chaotic behavior of the original differential equation \cite{alkahby}.
\end{defn}

\begin{defn}\label{def:NSFD}
A finite difference method is defined a NSFD scheme if at least one of the following conditions is satisfied \cite{mickens00}:
\begin{enumerate}
\item[I.] Nonlinear terms must be replaced by nonlocal discrete representations, i.e.,
\begin{align*}\xi^2\longrightarrow\xi_n\xi_{n+1}.\end{align*}
\item[II.] Denominator functions for the discrete representation must be nontrivial. The following replacement is then required: 
\begin{align*}\Delta t\longrightarrow\phi(\Delta t)+O(\Delta t^2),\end{align*}
where $\phi(\Delta t)$ is such that $0<\phi(\Delta t)<1$, for all $\Delta t>0$.
\end{enumerate}
\end{defn}
\begin{rem}
The NSFD scheme incorporates the principle of Dynamical Consistency.
\end{rem}
Other important rules to build discretization are:
\begin{itemize}
\item[- ] the order of the discrete derivative should be equal to the order of the corresponding derivatives of the differential equation;
\item[- ] special conditions that hold for the solutions of the differential equations should also hold for the solutions of the finite difference scheme;
\item[- ] the scheme should not introduce spurious solutions.
\end{itemize}

An important characteristic of dynamical systems, especially in those of biological interest, is that all solutions must remain nonnegative in order to maintain the problem well-posed, from biological and mathematical points of view. 

\begin{defn}\label{def:PESN}
A method that respects the NSFD rules and preserves the solution's positivity is called \textit{Positive and Elementary Stable Nonstandard} (PESN) method. 
\end{defn}

By applying these expedients, the discrete scheme will comply with the physical properties of the differential equations, without any restriction on the step size $\Delta t$.

\section{Hive bees \textit{vs} foragers population: the KMB model}

In this section the continuous-time model proposed in 2011 by Khoury, Myerscoug and Barron (KMB) is reported and briefly described; after, a discrete-time version is proposed.

\subsection{Continuous-time model}

The KMB model is based on hypothesis that a colony of honey bees is
schematizable as the sum of $h$ bees working in the hive and $f$
foragers bees working outside the hive, with a total number of workers $N=h+f$.
The system of differential equations proposed by the authors is:
\begin{equation}\label{eq:KMB}
\left\{
\begin{array}{l}
\frac{dh(t)}{dt}=L\frac{h(t)+f(t)}{w+h(t)+f(t)}-h(t)\bigg(\alpha-\sigma\frac{f(t)}{h(t)+f(t)}\bigg)  \\ \\

\frac{df(t)}{dt}=h(t)\bigg(\alpha-\sigma\frac{f(t)}{h(t)+f(t)}\bigg)-mf(t) \\ \\

h(0)\ge 0,\,f(0)\ge 0, \\
\end{array} 
\right.
\end{equation}
where the parameter $L$ is the maximum queen laying rate, $w$ represents the
brood mortality, $\sigma$ is the social inhibition and $\alpha$ is the maximum
rate at which hive bees will become foragers.
In the first differential equation of \eqref{eq:KMB}, the sign of
the right term depends upon the balance between a positive contribute, whose
magnitude is related to the \textit{eclosion rate}, and a negative
term, named  by the authors in the manuscript \textit{recruitment function}, due to
the hive bees recruited as foragers. The recruitment function is a balance between a
term that favors the transition to foragers and another that inhibits hive bees from transitioning to
foragers; with a such formulation, if the foragers number is high the
inhibition term prevents a further growing of $f$. \\Since it is not the purpose of this work to analyze the KMB model in all its details, the interested reader may find an exhaustive analysis in \cite{khoury2011}.

\subsection{Discrete-time model}

The differential equation \eqref{eq:KMB} can be discretized by
following the rules listed in Section \ref{sec:rulesNSFD}. The obtained difference equations are:

\begin{equation}\label{eq:KMB_timedisc}
\left\{
\begin{array}{l}
\frac{h_{n+1}-h_n}{\phi(\Delta t)}=L\frac{h_n+f_n}{w+h_n+f_n}-\alpha h_{n+1}+\sigma\frac{f_nh_n}{h_n+f_n}\\ \\

\frac{f_{n+1}-f_n}{\phi(\Delta t)}=h_n\bigg(\alpha-\sigma\frac{f_{n+1}}{h_n+f_n}\bigg)-mf_{n+1}\\ \\

h_0\ge 0, f_0\ge 0. \\
\end{array}
\right.
\end{equation}
That can be explicated respect to $h_{n+1}$ and $f_{n+1}$ as follows:
\begin{equation}\label{eq:KMB_timedisc_expl}
\left\{
\begin{array}{l}
h_{n+1}=\Phi_1(h_n,f_n)\\ \\

f_{n+1}=\Phi_2(h_n,f_n)\\ \\

h_0\ge 0, f_0\ge 0, \\
\end{array}
\right.
\end{equation}
in which
\begin{equation}\label{eq:KMB_timedisc_expl_1}
\left\{
\begin{array}{l}
\Phi_1(h_n,f_n)=\bigg(h_n\bigg(1+\frac{\sigma\phi(\Delta t)
f_n}{h_n+f_n}\bigg)+L\phi(\Delta
t)\bigg(1-\frac{w}{w+h_n+f_n}\bigg)\bigg)\frac{1}{1+\alpha\phi(\Delta t)}\\ \\

\Phi_2(h_n,f_n)=\frac{\big(f_n+h_n\big)\big(f_n+\alpha\phi(\Delta t) h_n\big)}{\big(1+m\phi(\Delta t)\big)f_n+\big(1+(m+\sigma)\phi(\Delta t)\big)h_n}\\ \\

h_0\ge 0, f_0\ge 0. \\
\end{array}
\right.
\end{equation}
To find the steady-states of \eqref{eq:KMB_timedisc_expl_1} the
condition \eqref{eq:fixed_point_disc} must be satisfied:
\begin{equation}\label{eq:KMB_steady_states}
\left\{
\begin{array}{l}
\Phi_1(h_n,f_n)=h_n\\ \\

\Phi_2(h_n,f_n)=f_n\\ \\

h_0\ge 0, f_0\ge 0. \\
\end{array}
\right.
\end{equation}
\begin{figure}
\centering
\includegraphics[width=0.95\textwidth]{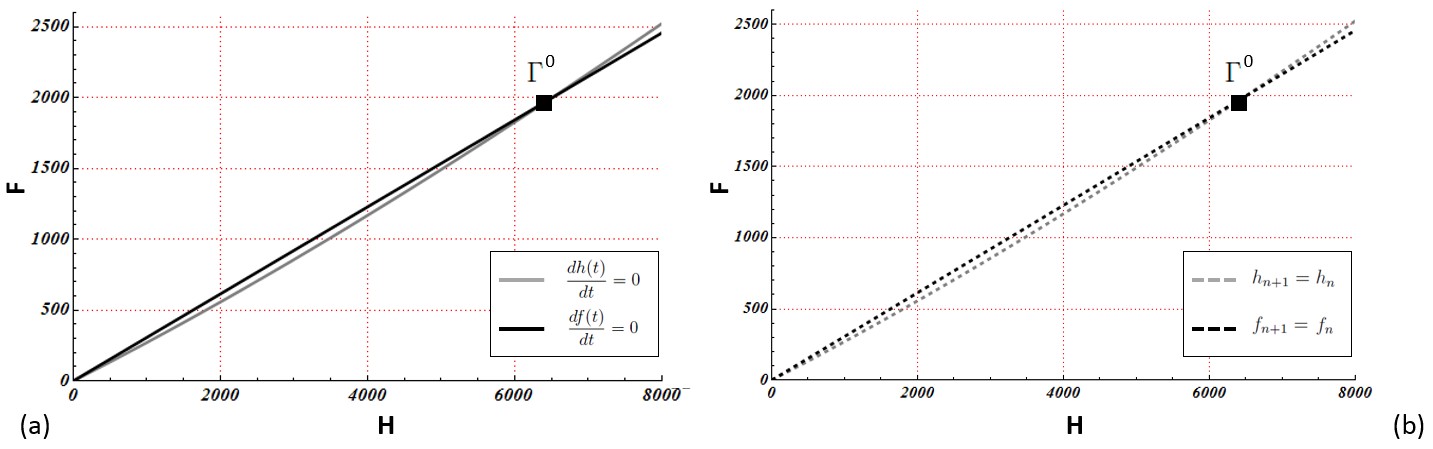}
\caption{\footnotesize{Comparison between phase-plane diagrams for continuous-time and discrete-time KMB models. (a) Zero-growth isoclines for system \eqref{eq:KMB} are represented as gray ($dh(t)/dt=0$) and black ($df(t)/dt=0$) lines. (b) Zero-growth isoclines of system \eqref{eq:KMB_timedisc}: black dotted line for the first equation and gray dotted line for the second equation. The steady-state is identified by the letter $\Gamma^0$. The values assigned at each parameter are: $L=2000$, $\alpha=0.25$, $\sigma=0.75$, $w=27000$ and $m=0.24$. The values assigned to the discretization parameters are: $\Delta t=0.1$ and $q=0.5$.}}\label{fig:phaseplane_KMB.jpg}
\end{figure}
The nontrivial form of $\phi(\Delta t)$, requested by the second rule of Definition \ref{def:NSFD}, is given by the following expression \cite{mickens00}:
\begin{equation}\label{eq:phi}
\phi(\Delta t)=\frac{1-e^{-q\Delta t}}{q}.
\end{equation}
Introducing $\Omega=\bigcup_{\tilde\xi\in\Gamma_d}\sigma(\tilde J)$,
the optimal value of $q$ must respect the condition
\begin{equation}\label{eq:condition}
q\ge\max_{\Omega}\Big\{\frac{\lambda^2}{2|\Re(\lambda)|}\Big\}\,\,\,\rm if\,\,\,  \Re(\lambda)\neq 0\,\, for    \,\,\,\lambda\in\Omega.
\end{equation}
By assuming $L=2000$, $\alpha=0.25$, $\sigma=0.75$, $w=27000$ and $m=0.24$, the same values adopted in \cite{khoury2011}, the zero-growth isoclines in continuous and in discrete cases are represented
respectively in \figurename~\ref{fig:phaseplane_KMB.jpg}\textit{a} and \textit{b}. In both cases, they intersect in the point $\Gamma^0=(H^0,F^0)\simeq(6470,1988)$: this means that $\Gamma^0$ constitutes an
equilibrium point of both the systems \eqref{eq:KMB}
and \eqref{eq:KMB_timedisc}. 

\begin{figure}
\centering
\includegraphics[width=0.95\textwidth]{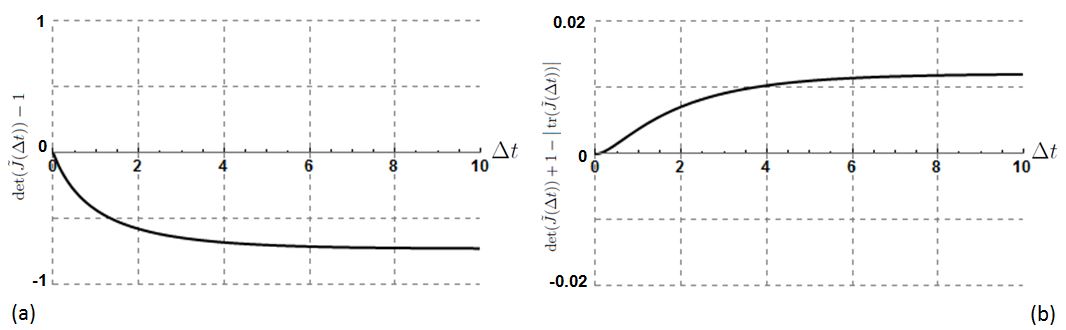}
\caption{\footnotesize{Jury condition validation for NSFD scheme. (a) Plot of $\det(\tilde J(\Delta t))-1$ versus the time step $\Delta t$. (b) Plot of $\det(\tilde J(\Delta t))+1-\big|\operatorname{tr}(\tilde J(\Delta t))\big|$ versus $\Delta t$.}}\label{fig:jury}
\end{figure}

To evaluate the stability of $\Gamma_0$ in continuous-time, the definition \ref{def:stab_cont} must be taken into account. In particular, the set of eigenvalues calculated in point $\Gamma^0$ is $\sigma(J_F(H^0,F^0))\simeq\{-0.83,-0.014\}$; since they
are both negatives, is possible to conclude that $\Gamma^0$ is
asymptotically stable. The magnitude of the eigenvalues implies a choice of
$q\ge 0.41$, then a $q=0.5$ was used for the numerical calculations described in the following sections.

In the discrete-time case, in order to evaluate the stability of $\Gamma_0$, is sufficient to consider the inequality \eqref{eq:Jury} and rewrite it making explicit the dependence on $\Delta t$ in order to control if the step size affects or not the stability of $\Gamma^0$:

\begin{equation}\label{eq:det-1}
\det(\tilde J(\Delta t))-1<0,
\end{equation}
\begin{equation}\label{eq:det+1-tr}
\det(\tilde J(\Delta t))+1-\big|\operatorname{tr}(\tilde J(\Delta t))\big|>0.
\end{equation}
The \figurename~\ref{fig:jury}\textit{a} and \figurename~\ref{fig:jury}\textit{b} show the left terms of inequalities expressed respectively in \eqref{eq:det-1} and in \eqref{eq:det+1-tr}: it is possible to note that both the inequalities \eqref{eq:det-1} and \eqref{eq:det+1-tr} are respected up to $\Delta t=10$, then, in this time-step interval, $\Gamma^0$ is asymptotically stable as found in continuous case.

\section{NAKMB: the proposed nonautonomous KMB model}

In this section a nonautonomous version of the KMB model is
proposed, in which the seasonal effects are investigated by introducing a
time-dependent formulation for some of the parameters involved in KMB
model. Also, differently from the original KMB model, in addition to the foragers bees death rate, a nonzero death rate for hive bees was assumed in order to reproduce a possible depopulation related to inadequate nutrition, brood disease or to the presence of Varroa mites and viruses, among the main factors \cite{brown2013}.
  
\subsection{Time-dependent parameters}

A healthy colony of honey bees regulates the temperature of the nest $T^{N}$ using heating and cooling systems \cite{cramp2008}, to compensate for the
variations of the environmental temperature $T^E$. In \cite{hess1926,himmer1927} was highlighted that, in order to grant the proper rearing of brood, an optimal value for the nest temperature $T^O$ exists, comprised in the narrow range of $32$-$36\,^{\circ}C$, with a mean of $34.5\,^{\circ}C$. In case of necessity, in order to cool the hive during the warmer months, the worker bees start fanning,
evaporate water by tongue lashing or spread droplets of water on the brood \cite{becher2009}; conversely, to heat the nest, the bees form a cluster
clinging to each other. The role of the temperature, that of the nest in direct way and the environmental one in a roundabout way, is then of the utmost importance and it deserves to be taken into account in a model that aims to describe how the population evolves in time.

Below, the fundamental hypotheses at the base of the NAKMB model are discussed and the mathematical formulation is given:

\begin{itemize} 
\item[1.] The queen laying rate $L$ is influenced by the nest temperature. This assumption is justified by the evidence that a low deposition rate is linked to a $T^N$ substantially greater or smaller than $T^O$ \cite{dunham1930}. To reproduce this behavior the proposed formula is:
\begin{align}\label{eq:laying_rate}
L(t)=\frac{L_0(\Gamma/2)^2}{(T^N(t)-T^O)^2+(\Gamma/2)^2},
\end{align}
where $L_0$ is the maximum laying rate. The function defined in \eqref{eq:laying_rate} is a Lorentzian curve characterized by a width $\Gamma$, multiplied by a factor such that, if $T^N=T^O$, then $L_{max}=L_0$. The $\Gamma$ represents the ability to withstand deviations of the nest temperature from the optimal value. 

\item[2.] The nest temperature is maintained around the optimal value
if the total number of individuals in a colony $N$ overcomes a critical threshold $N^T$; if the number of individuals falls below that threshold the colony may fail to incubate brood or maintain the optimum temperature in the nest \cite{rosenkranz2008,khoury2013}. To capture this effect the following formulation is proposed:
\begin{align}\label{eq:Tnest}
T^N(t)=\frac{T^E(t)}{1+\big(\frac{f+h}{N^T}\big)^2}+\frac{T^O}{1+\big(\frac{N^T}{f+h}\big)^2}
\end{align}

\item[3.] The environmental temperature $T^E$ is supposed to have sinusoidal behavior with annual periodicity:
\begin{align}\label{eq:Tenv}
T^E(t)=\theta_0+\theta_1\sin(\Omega t)+\theta_2\cos(\Omega t),
\end{align}
in which $\Omega=2\pi/\tau$, $\tau$ is the period. The coefficients $\theta_0$, $\theta_1$ and $\theta_2$ can be estimated by least-square fitting of the environmental temperature measured, for example, by a weather station.

\item[4.] The recruitment function, due to hive bees recruited as
foragers, unlike the KMB model, is supposed to be influenced by the season. To meet this need, it can be multiplied by a coefficient
\begin{align}\label{eq:eta}
\eta(t)=\eta_0+\eta_1\sin(\Omega t)+\eta_2\cos(\Omega t).
\end{align}
In particular, since the foraging activity stops during the winter
and is maximum during the spring, $\eta(t)$ must have zero-phase delay respect to $T^E(t)$ (maxima and minima coinciding) and respect $0\le\eta(t)\le 1$. To satisfy these two requirements the coefficients in \eqref{eq:eta} must be linked to $\theta_1$ and $\theta_2$ as follows (see the proof in the Appendix):
\begin{equation}\label{eq:sol_eta}
\eta_0=\frac{1}{2},\,\,\eta_1=-\frac{\theta_1\theta_2}{2\big(\theta_1^2+\theta_2^2\big)}\sqrt{1+\bigg(\frac{\theta_1}{\theta_2}\bigg)^2},\,\,\eta_2=-\frac{\theta_2^2}{2\big(\theta_1^2+\theta_2^2\big)}\sqrt{1+\bigg(\frac{\theta_1}{\theta_2}\bigg)^2}
\end{equation}
\end{itemize}

\begin{figure}[!ht]
\includegraphics[width=\textwidth]{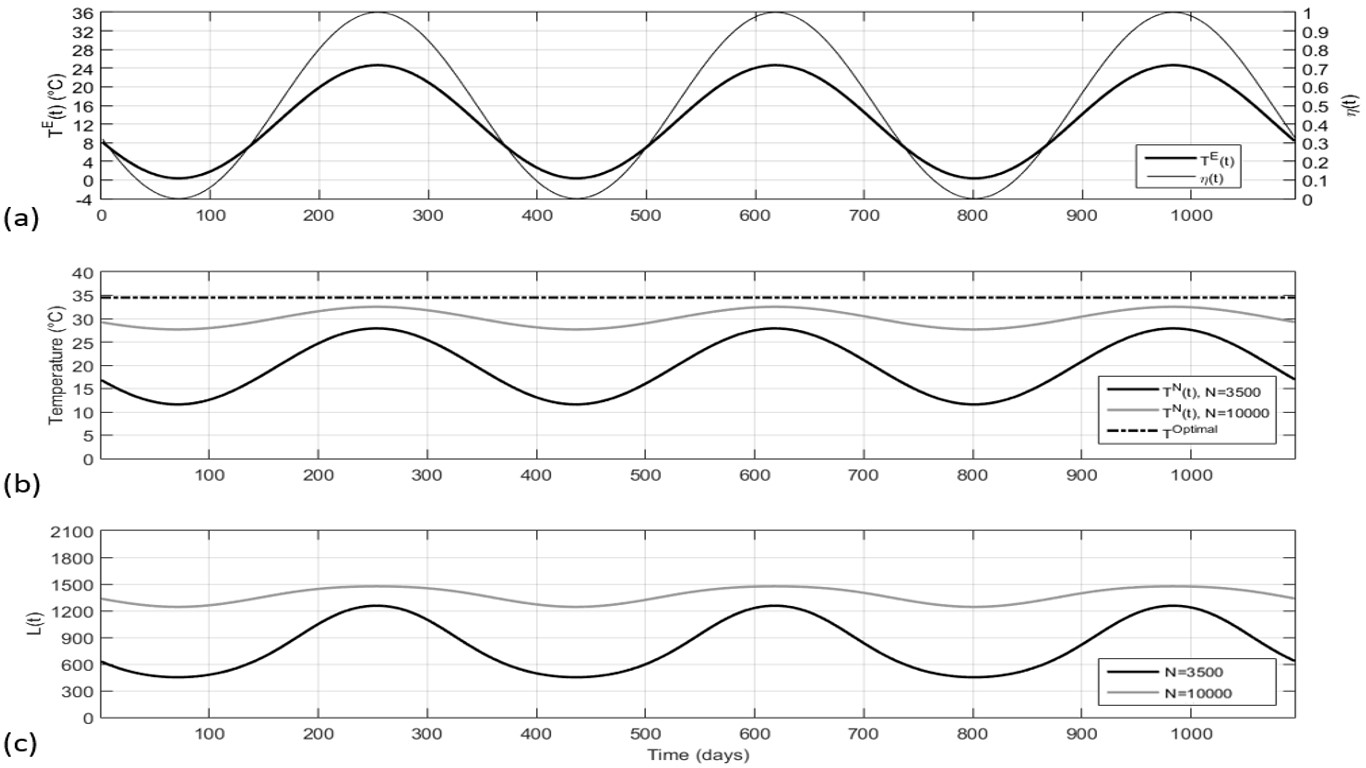}
\caption{\footnotesize{Temporal evolution of the environmental temperature $T^E(t)$, recruiting function coefficient $\eta(t)$, nest temperature $T^N(t)$ and queen laying rate $L(t)$ over a period of 2 years. (a) $T^E(t)$ of \eqref{eq:Tenv} with
$\theta_0=12.51\,^{\circ}{\rm C}$, $\theta_1=-11.41\,^{\circ}{\rm C}$, $\theta_2=-4.16\,^{\circ}{\rm C}$ and
$\Omega=2\pi/\tau$, with period $\tau=365$ days (thick line); $\eta(t)$ of \eqref{eq:eta} with $\eta_0=0.5$, $\eta_1=-0.47$ and $\eta_2=-0.17$ (thin line). (b) $T^N(t)$ of \eqref{eq:Tnest} for $N_0=3500$ (black) and $N_0=10000$ individuals (gray) with $N^T=5000$ individuals; $T^O=34.5\,^{\circ}{\rm C}$ (black dashed line). (c) $L(t)$ of \eqref{eq:laying_rate} for $N_0=h_0+f_0=3500$ (black) and $N_0=10000$ (gray), with $T^O=34.5\,^{\circ}{\rm C}$ and $\Gamma=30$. }}\label{fig:parameters_NAKMB}
\end{figure}

In Figures \ref{fig:parameters_NAKMB}\textit{a}, \textit{b}
and \textit{c} are shown respectively the temporal evolutions of
$T^E(t)$, $\eta(t)$, $T^N(t)$ and $L(t)$ over a period of 2 years, keeping fixed the number of individuals of each population ($h=h_0$, $f=f_0$ and $N=N_0=h_0+f_0$). In particular, in \figurename~\ref{fig:parameters_NAKMB}\textit{a} is
shown the $T^E(t)$, whose coefficients $\theta_0$, $\theta_1$ and
$\theta_2$ were evaluated by nonlinear least-square fitting of
the annual variation of temperature registered by the meteorological
station located in Osnago, Italy (lat: $45.68^{\circ}$, long:
$9.38^{\circ}$) during the year 2012 \cite{cml}. Superimposed on
$T^E(t)$ there is $\eta(t)$, whose coefficients $\eta_0$, $\eta_1$ and
$\eta_2$ are calculated by using \eqref{eq:sol_eta}. By
setting the threshold population at $5000$ individuals \cite{xia2016}, the nest
temperature is calculated and shown
in \figurename~\ref{fig:parameters_NAKMB}\textit{b} for two values of
the total populations $N=3500$ and $N=10000$, respectively below and above the critical threshold. In \figurename~\ref{fig:parameters_NAKMB}\textit{c} is represented the queen laying rate for $N=3500$ and $N=10000$.

From the \figurename~\ref{fig:parameters_NAKMB}\textit{b}, it is possible to see that the nest temperature is maintained close to the optimal value if the population is greater than the critical threshold, otherwise is greatly influenced by the environmental temperature. Consequently, being $L(t)$ dependent on $T^N(t)$, and being $T^N(t)$ linked to the total number of individuals in a colony, results that $L(t)$ is sensitively influenced by the number of individuals in a colony. 

\subsection{Continuous-time NAKMB model}

The system of autonomous nonlinear differential equations at the base of the KMB
model is modified as follows:

\begin{equation}\label{eq:NAKMB}
\left\{
\begin{array}{l}
\frac{dh(t)}{dt}=L(t)\frac{h(t)+f(t)}{w+h(t)+f(t)}-\eta(t)h(t)\bigg(\alpha-\sigma\frac{f(t)}{h(t)+f(t)}\bigg)-\mu h(t)  \\ \\
\frac{df(t)}{dt}=\eta(t) h(t)\bigg(\alpha-\sigma\frac{f(t)}{h(t)+f(t)}\bigg)-mf(t) \\ \\
h(0)\ge 0,\,f(0)\ge 0, \\
\end{array} 
\right.
\end{equation}
where $L(t)$ and $\eta(t)$ are expressed respectively in \eqref{eq:laying_rate} and in \eqref{eq:eta}. In the right term of the
first equation $\mu$ represents the death rate factor afflicting the
hive bees. Since the laying rate and the recruiting function
coefficients were assumed explicitly depending on time, the model defined
in \eqref{eq:NAKMB} is nonautonomous.

\subsection{Discrete-time NAKMB model: a nonstandard formulation}

The system \eqref{eq:NAKMB} is now formulated in discrete-time
domain by taking advantage of the NSFD scheme. The proposed formulation is:
\begin{equation}\label{eq:NAKMB_timedisc}
\left\{
\begin{array}{l}
\frac{h_{n+1}-h_n}{\phi(\Delta t)}=L_n\frac{h_n+f_n}{w+h_n+f_n}-\alpha \eta_n h_{n+1}+\sigma\eta_n\frac{f_nh_n}{h_n+f_n}\\ \\

\frac{f_{n+1}-f_n}{\phi(\Delta t)}=\eta_nh_n\bigg(\alpha-\sigma\frac{f_{n+1}}{h_n+f_n}\bigg)-mf_{n+1}\\ \\

h_0\ge 0, f_0\ge 0, \\
\end{array}
\right.
\end{equation}
in which
\begin{align}\label{eq:laying_rate_disc}
L_n=\frac{L_0(\Gamma/2)^2}{(T^N_n-T^O)^2+(\Gamma/2)^2},
\end{align}
\begin{align}\label{eq:Tnest_disc}
T^N_n=\frac{T^E_n}{1+\big(\frac{f_n+h_n}{N^T}\big)^2}+\frac{T^O}{1+\big(\frac{N^T}{f_n+h_n}\big)^2},
\end{align}
\begin{align}\label{eq:Tenv_disc}
T^E_n=\theta_0+\theta_1\sin(\Omega n)+\theta_2\cos(\Omega n),
\end{align}
\begin{align}\label{eq:eta_disc}
\eta_n=\eta_0+\eta_1\sin(\Omega n)+\eta_2\cos(\Omega n).
\end{align}
With simple algebraic manipulation, the explicit form of $h_{n+1}$ and $f_{n+1}$ is obtained:
\begin{equation}\label{eq:NAKMB_disc_expl}
\left\{
\begin{array}{l}
h_{n+1}=\bigg(h_n+L_n\frac{h_n+f_n}{w+h_n+f_n}+\sigma\phi(\Delta t)\frac{\eta_nf_nh_n}{f_n+h_n}\bigg)\big(1+\phi(\Delta t)(\mu+\alpha\eta_n)\big)^{-1}\\ \\

f_{n+1}=\frac{\big(f_n+\alpha\phi(\Delta t)\eta_nh_n\big)}{\sigma\phi(\Delta t)\big(1+m\phi(\Delta t)\big)\eta_nh_n}\\ \\

h_0\ge 0, f_0\ge 0. \\
\end{array}
\right.
\end{equation}
The choice of the NSFD method has left the chance to customize the discretization of \eqref{eq:NAKMB} in order to verify the positivity condition for $h_{n+1}$ and $f_{n+1}$. In fact, since all the parameters are nonnegative, the condition $h_{n+1},f_{n+1}\ge 0$ is verified for all $n\in\mathbb{Z}^+$, then the NAKMB respects the PESN criterion formalized in Definition \ref{def:PESN}. 

\subsection{Phase-plane analysis}

\begin{figure}[!ht]
\centering
\includegraphics[width=\textwidth]{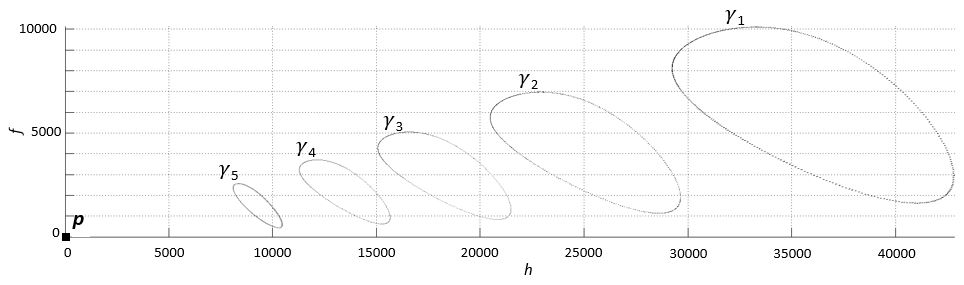}
\caption{\footnotesize{Phase-plane of $h$ and $f$ expressed in \eqref{eq:NAKMB_disc_expl}. The trajectories start from $2000\le h_0\le 20000$, $500\le f_0\le 8000$. The steady states, labelled with $\gamma_1\div\gamma_5$, and the equilibrium point at $(h,f)=(0,0)$, labelled with $p$ and highlighted with a black square, are obtained with different values of $\mu$: $0$, $0.015$, $0.03$, $0.045$, $0.06$ and $0.07$. The other parameters are: $L_0=2000$, $\alpha=0.25$, $\sigma=0.75$, $\omega=10000$, $m=0.24$, $N^T=5000$, $T^O=34.5\,^{\circ}{\rm C}$, $\theta_0=12.51\,^{\circ}{\rm C}$, $\theta_1=-11.41\,^{\circ}{\rm C}$, $\theta_2=-4.16\,^{\circ}{\rm C}$, $\Gamma=30$ and $\Omega=2\pi/\tau$, with period $\tau=365$ days. Number of iterations: $3\cdot 10^4$.}}\label{fig:phase_plane}
\end{figure}

\begin{figure}[!ht]
\centering
\includegraphics[width=\textwidth]{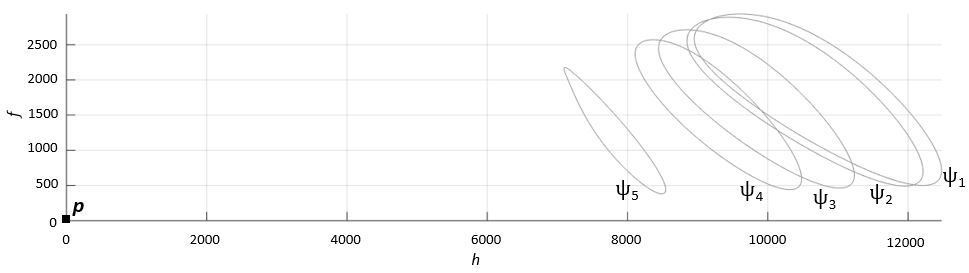}
\caption{\footnotesize{Phase-plane of $h$ and $f$ expressed in \eqref{eq:NAKMB_disc_expl}. The trajectories start from $2000\le
h_0\le 20000$ and $500\le f_0\le 8000$ and increase with a step of $500$ individuals. The steady states, labelled
with $\psi_1\div\psi_5$, and the equilibrium point at $(h,f)=(0,0)$, labelled with $p$ and highlighted by a black square, were obtained with different values of $\Gamma$: $100$, $60$, $35$, $30$ and $25$.  The number of iterations and the values attributed to the parameters are the same used for computation of \figurename~\ref{fig:phase_plane}, except for $\mu=0.06$.}}\label{fig:phase_plane_gamma}
\end{figure}

The phase-plane of the discrete NAKMB model is calculated, by assuming the initial conditions $2000\le h_0\le 20000$, $500\le f_0\le 8000$ with an increment of $500$ individuals. The main purpose is to evaluate the effects of the parameters $\mu$ and $\Gamma$ on the asymptotical behavior of the system.

In \figurename~\ref{fig:phase_plane} the phase-plane of NAKMB model is represented for different values of the parameter $\mu$: $0$, $0.015$, $0.03$, $0.045$, $0.06$ and $0.07$. It is possible to recognize five closed loops, labelled with $\gamma_1$-$\gamma_5$ and an equilibrium point labelled with $p$, located at $(h,f)=(0,0)$ and highlighted by a black square. The closed loops $\gamma_1$, $\gamma_2$ and $\gamma_3$, classifiable as limit cycles, appear respectively assuming $\mu=0,\,0.015,\,0.03$; by choosing $\mu=0.045,\,0.06$ arise respectively the limit cycles $\gamma_4$ and $\gamma_5$ and the equilibrium point $p$; $\mu=0.07$ leads the system to have only the equilibrium point $p$. With an accurate analysis it was possible to verify that there exists a threshold value $\tilde\mu \simeq 0.038$ above which the system exhibits the two aforementioned asymptotic behaviors: depending on the initial conditions, the population can settle in a limit cycle or definitely decline; similarly, it is possible to show that assuming $\mu>\tilde\mu'\simeq 0.066$ the system goes toward the equilibrium point $p$, regardless to the initial conditions. The autonomous model studied by Brown \cite{brown2013}, in which is also considered $\mu\neq 0$, a critical threshold has been established at $\tilde\mu'_B=0.12234969$ above which occurs the collapse of the population, by assuming a constant laying rate $L=1500$. Then, in NAKMB model, unlike in the autonomous Brown's model, values of $\tilde\mu'\le\mu\le\tilde\mu'_B$ lead to the collapse of the colony. By looking at the \figurename~\ref{fig:phase_plane} it is also remarkable that the range of population in which are enclosed the limit cycles decreases by increasing the hive bees death rate $\mu$, as it is reasonable to be.   

\begin{figure}[!ht]
\centering
\includegraphics[width=0.9\textwidth]{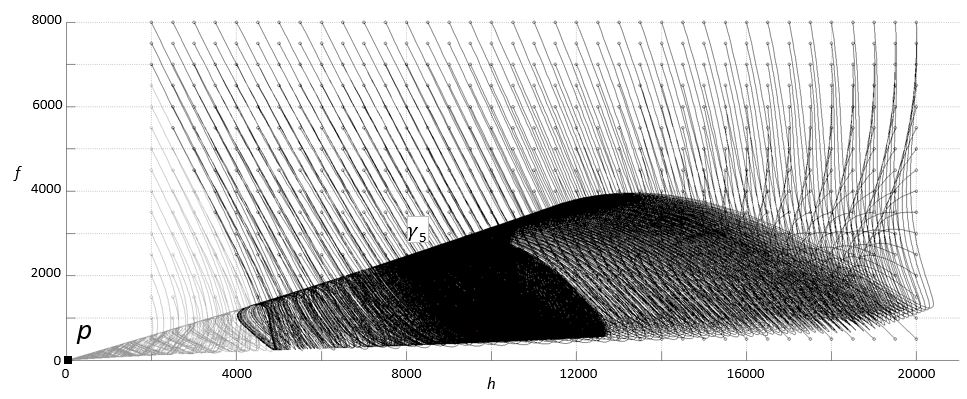}
\caption{\footnotesize{Phase-portrait of $h$ and $f$ expressed in \eqref{eq:NAKMB_disc_expl} using the same parameters values used for computation of phase-plane in \figurename~\ref{fig:phase_plane} and $\mu=0.06$. The trajectories start from the initial points $2000\le h_0\le 20000$, $500\le f_0\le 8000$ and increase with a step of $500$ individuals. Number of iterations: $3\cdot 10^4$. The trajectories ending in $p$ are colored in gray and the ones ending in limit cycle $\gamma_5$ in black.}}\label{fig:phase_portrait}
\end{figure}

By fixing $\mu=0.06$, as in \cite{brown2013}, the phase-plane of NAKMB model is calculated and represented in \figurename~\ref{fig:phase_plane_gamma} for different values of the parameter $\Gamma$: $100$, $60$, $35$, $30$ and $25$. The limit cycles $\psi_1$ and $\psi_2$ appear respectively assuming $\Gamma=100$ and $60$; the limit cycles $\psi_3$, $\psi_4$ and $\psi_5$, together with the equilibrium point $p$ characterized by $(h,f)=(0,0)$, appear assuming respectively $\Gamma=35,\,30$ and $25$. A detailed analysis showed that the threshold between these two behaviors is $\tilde\Gamma\simeq 39.5$. Similarly, it exists a second threshold $\tilde\Gamma'\simeq 25.9$, so that if $\Gamma<\tilde\Gamma'$ is got only the final state $p$; for this reason, the trajectory calculated assuming $\Gamma=25$ leads to $p$. This means that the larger is $\Gamma$, the better is the ability of a colony to survive to large variations of the nest temperature; smaller values lead towards a collapse of the colony. Then, an appropriate value of $\Gamma$,  presumably in the range $\tilde\Gamma'<\Gamma<\tilde\Gamma$, must be chosen in order to have two possible asymptotic states depending on the initial conditions: a limit cycle whose populations $h,\, f$ settle around ecologically plausible values and an equilibrium point representing the collapse of the colony. 

In order to delineate the geometry of the basins of attraction, characterized by the set of initial conditions leading to a certain attractor, the phase portrait of \eqref{eq:NAKMB_disc_expl} is calculated and represented in \figurename~\ref{fig:phase_portrait}, by fixing $\mu=0.06$ and $\Gamma=30$. The trajectories colored in gray are those driving the system to the collapse and the black trajectories lead to the limit cycle $\gamma_5$. It is easily verifiable that the two basins of attraction are separated by a straight line of equation $f(h)\simeq -1.8462 h+9915$.  

\subsection{Toroidal phase-space representation}

The temporal evolution of the NAKMB dynamical system through the phase-portrait of \figurename~\ref{fig:phase_portrait} is rather complicated to be visually interpreted, since the trajectories intersect between them. The NAKMB model is, in fact, a nonautonomous model, in which the parameters $L(t)$ and $\eta(t)$ are $\tau$-periodic, then
$L(t)=L(t+\tau)$ and $\eta(t)=\eta(t+\tau)$. In discrete domain, the periodicity condition corresponds to $L_n=L_{n+\nu}$ and $\eta_n=\eta_{n+\nu}$, in which $\nu=2\pi/\Omega$ is the period. Following the approach of \eqref{eq:differential_equation_na_1}, the behavior of the system of \eqref{eq:NAKMB_disc_expl} can be investigated by
introducing the angular variable $\theta$ so that replacing $\theta+2\pi m$ for $\theta$, $m\in\mathbb{Z}^+$, the system is unchanged.  

A particularly suitable alternative to visualize and analyze the trajectories of periodic nonautonomous dynamical systems is given by the toroidal phase-space \cite{jordan2007}. As introduced in Subsection \ref{subsec:poincare}, it allows an expeditious computation of the Poincar\'e map. The steps carried out to construct this representation are the following:

\begin{itemize}

\item[I.] Let $D_0$ the domain composed by the initial conditions and $D_1$ the rectangular domain enclosing the trajectories starting from $D_0$ for all $t$, or $n$ in discrete case ($D_0\subseteq D_1$). Suppose that $D_1$ is defined by $h_a\le h\le h_b,\,f_a\le f\le f_b$, the new coordinates are calculated as follows:
\begin{align}\label{eq:norm}
h'=\frac{h-h_a}{h_b-h_a},\,\,f'=\frac{f-f_a}{f_b-f_a},
\end{align}
in which $h_b\neq h_a$ and $f_b\neq f_a$. A such transformation ensures $0\le h'\le 1$ and $0\le f'\le 1$.

\item[II.] $h'$-$f'$ phase-plane is mapped into a $h^*$-$f^*$ plane by applying the diametrical transformation \cite{jordan2007}: 
\begin{align}\label{eq:diametrical}
h^*=\frac{h'}{\sqrt{1+r^2}},\,\,f^*=\frac{f'}{\sqrt{1+r^2}},
\end{align}
in which 
\begin{align}
r=\sqrt{h'^2+f'^2}.
\end{align}
This projection allows visualizing the whole phase-plane in a circumference of unitary radius.
\end{itemize}

Although the step I is not strictly required, with $h,f$ varying in the range $0$-$10^4$ individuals, the diametrical projection applied directly to $h$ and $f$ would have led to trajectories squeezed around the outer surface of the manifold $\mathbb{R}^{n}\times S^1$; the transformation performed in step I allows to distribute the trajectories in a larger portion of the manifold to improve the visualization.

\begin{figure}[!ht]
\centering
\includegraphics[width=0.9\textwidth]{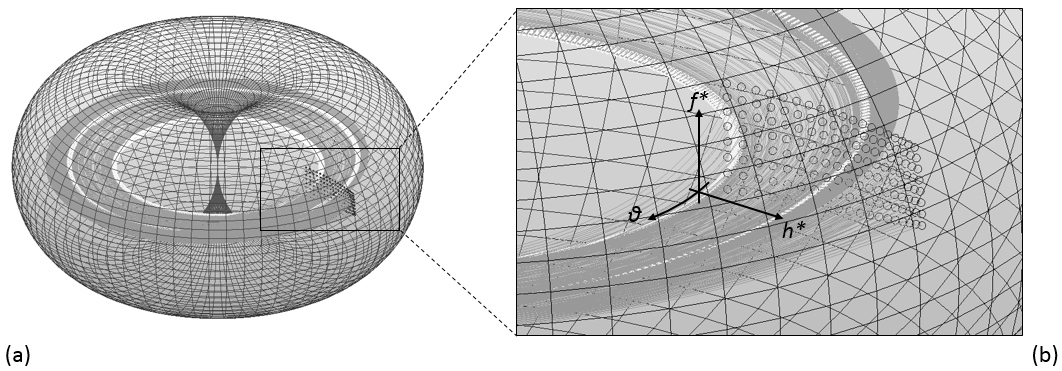}
\caption{\footnotesize{(a) Toroidal phase-space of the discrete NAKMB model of \eqref{eq:NAKMB_disc_expl}. (b) Zoom on the initial conditions, represented by black circles. The trajectories are represented in gray far from equilibrium and in white approaching to the equilibrium. The torus has unitary major and minor radii.}}\label{fig:toro}
\end{figure}

The toroidal phase-space is shown in \figurename~\ref{fig:toro}\textit{a} adopting the same parameter values used to compute the \figurename~\ref{fig:phase_portrait}. The orbits start from the points indicated with the black circles located at $\theta=0$ and back after $2\pi m$, $m\in\mathbb{Z^+}$; the trajectories are colored in gray far from equilibrium and in white approaching to the equilibrium. The inner white closed path corresponds to the equilibrium point \textit{p} characterized by $(h,f)=(0,0)$ in the standard phase-plane, and the outer one to the limit cycle $\gamma_5$. 

\begin{figure}[!ht]
\centering
\includegraphics[width=0.9\textwidth]{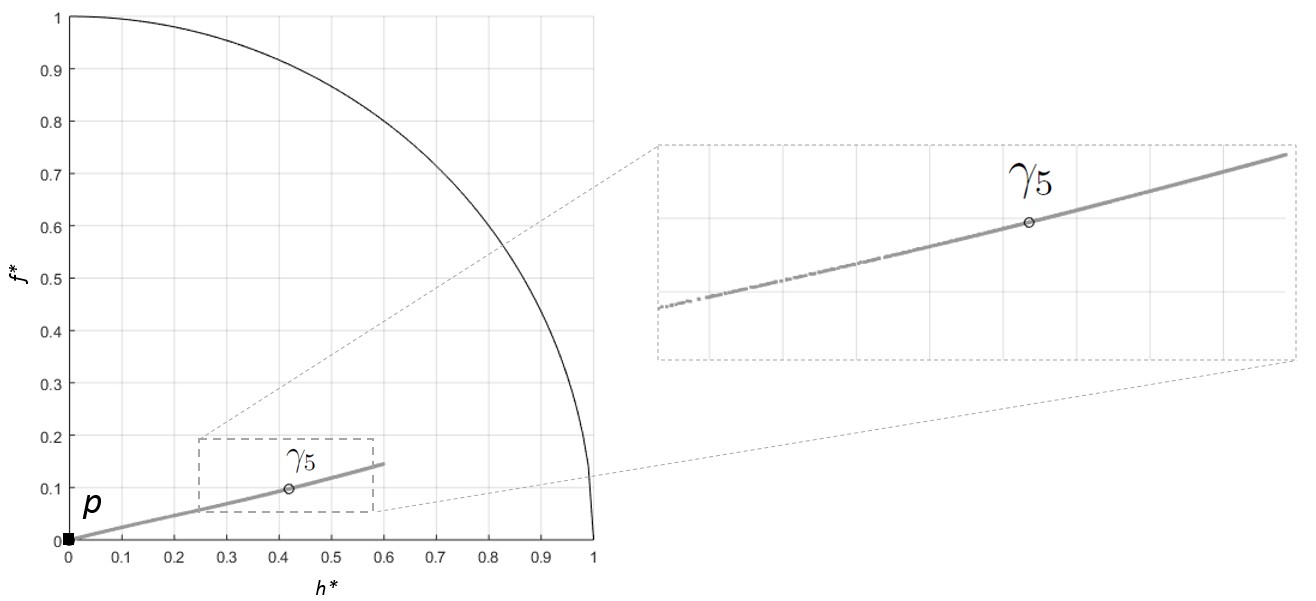}
\caption{\footnotesize{Poincar\'e map extracted as a slice at $\theta=\pi/2$ of the torus in \figurename~\ref{fig:toro}; only the quadrant $h^*,f^*\ge 0$ is represented.}}\label{fig:poincaremap}
\end{figure}

The toroidal phase-space computation facilitates the extraction of the Poincar\'e map since it represents a section of the torus at a particular value of $\theta$, which is a disc of unitary radius. By fixing $\theta=\pi/2$ the Poincar\'e map of \figurename~\ref{fig:poincaremap} is obtained; note that only the quadrant $h^*,f^*\ge 0$ is represented. The equilibrium point $\textit{p}$ and the limit cycle $\gamma_5$ are highlighted respectively by a square and a circle. By looking at the zoomed window of \figurename~\ref{fig:poincaremap}, it is possible to note the progressive densification of the points approaching $\gamma_5$, as well as in phase-portrait of \figurename~\ref{fig:phase_portrait} there is a densification of the intersecting paths approaching $\gamma_5$. 

\section{Conclusions}

In this paper the continuous-time model proposed by Khoury \textit{et al.} \cite{khoury2011} (KMB) has been considered and formulated in discrete-time domain by adopting the Nonstandard Finite Difference (NSFD) scheme. As shown by the numerical simulations, the NSFD scheme well reproduces the behavior of the continous-time KMB model, in terms of steady-states and their stability properties. 

Subsequently, a nonautonomous version of the KMB model was proposed (NAKMB), in which the seasonal effects were introduced by means a time-dependent formulation for the queen laying rate and the eclosion rate coefficient. In particular, the formulation proposed for the queen laying rate was meant to reproduce the experimental evidence that a reduced number of individuals in a colony, below a critical threshold, can hardly be able to maintain the optimum temperature in the nest, especially in the colder periods, and this could lead the entire colony toward the collapse. Taking advantages of the NSFD scheme, the NAKMB model was studied via a numerical approach in standard phase-plane and in toroidal phase-space. The system showed a sensitive dependence of the steady-states on parameters $\mu$ and $\Gamma$, respectively the hive bees death rate and the parameter introduced in the present study representing the ability of a colony to bear large variations of the nest temperature from the optimal value. In particular, small values of $\mu$ lead the population toward a limit cycle regardless of the initial conditions; by increasing $\mu$ the colony can experience a collapse or stabilizes in a limit cycle depending on the initial population; high values of $\mu$ bring the system toward the collapse, regardless of the initial population. The same three scenarios are obtained by varying $\Gamma$ from high to low values.
An interesting feature of the NAKMB model is that, by assuming a moderate death rate for the foragers bees ($m=0.24$) and with an appropriate choice of the parameters, it admits two steady-states; unlike the KMB model that predicts only one equilibrium point at nonzero populations, regardless of the initial population. This features of the NAKMB model seems to be reasonable since a colony counting a number of individuals below the critical threshold encounter more difficulties to develop respect to a large colony and its collapse is a possible scenario.

\appendix

\section{Recruiting function coefficient}

The formula proposed for the recruiting function coefficient is:
\begin{align}\label{eq_app:eta}
\eta(t)=\eta_0+\eta_1\sin(\Omega t)+\eta_2\cos(\Omega t).
\end{align}
In this section the explicit formulation of $\eta_0$, $\eta_1$ and
$\eta_2$ coefficients is derived. The requisites that $\eta(t)$ must respect are:
\begin{itemize}
\item[I.] Zero-phase delay respect to $T^E(t)$ of \eqref{eq:Tenv}.

Since
\begin{align}
\frac{dT^E(t)}{dt}=\Omega \big(\theta_1\cos(\Omega t)-\theta_2\sin(\Omega t)\big),
\end{align}
for the extreme points of $T^E(t)$:
\begin{align}
\frac{dT^E(t)}{dt}=0\,\, \iff t=\hat t=\frac{1}{\Omega}\bigg(\arctan\bigg(\frac{\theta_1}{\theta_2}\bigg)+k\pi\bigg),\,\,k\in\mathbb{Z}.
\end{align}
Similarly, for the extreme points of $\eta(t)$:
\begin{align}
\frac{d\eta^E(t)}{dt}=0\,\, \iff t=\tilde t=\frac{1}{\Omega}\bigg(\arctan\bigg(\frac{\eta_1}{\eta_2}\bigg)+k\pi\bigg),\,\,k\in\mathbb{Z}.
\end{align}
In order to have zero-phase delay between $\eta(t)$ and $T^E(t)$,
is required that $\hat t=\tilde t$, then the relation 
\begin{align}\label{eq:rel_par}
\eta_1 \theta_2=\eta_2 \theta_1
\end{align}
must be respected. 

After a simple algebraic manipulation, the second derivative becomes:
\begin{align}\label{eq:der2eta}
\frac{d^2\eta(t)}{dt^2}\bigg{|}_{t=\tilde t}=(-1)^k\frac{\eta_1\Omega^2\theta_2}{\theta_1}\bigg(1+\bigg(\frac{\theta_1}{\theta_2}\bigg)^2\bigg)^{1/2}.
\end{align}
By looking at \eqref{eq:der2eta}, it is possible to conclude that if $k$ is even, then $\tilde t$ is a relative minimum; if $k$ is odd, then $\tilde t$ is a maximum.

\item[II.] $0\le\eta(t)\le 1$.

From \eqref{eq:rel_par}, $\eta_2 =\eta_1\theta_2/\theta_1$ may be substituted in \eqref{eq_app:eta}. Then, the coefficients $\eta_0$ and $\eta_1$ can be found by solving the system:
\begin{equation}\label{eq:sist_eta}
\left\{
\begin{array}{l}
\eta_0+\eta_1\bigg(\sin(\Omega t)+\frac{\theta_2}{\theta_1}\cos(\Omega t)\eta(t)\bigg)=1,\,\,\forall t\in S^M \\

\eta_0+\eta_1\bigg(\sin(\Omega t)+\frac{\theta_2}{\theta_1}\cos(\Omega t)\eta(t)\bigg)=0,\,\,\forall t\in S^m, \\
\end{array}
\right.
\end{equation}
where $S^M$ is the set of the relative maximum and $S^m$ of the relative minimum. The solutions of \eqref{eq:sist_eta} are:
\begin{equation}\label{eq_app:sol_eta}
\eta_0=\frac{1}{2},\,\,\eta_1=-\frac{\theta_1\theta_2}{2\big(\theta_1^2+\theta_2^2\big)}\sqrt{1+\bigg(\frac{\theta_1}{\theta_2}\bigg)^2},
\end{equation}
then,
\begin{align}
\eta_2=-\frac{\theta_2^2}{2\big(\theta_1^2+\theta_2^2\big)}\sqrt{1+\bigg(\frac{\theta_1}{T^b}\bigg)^2}.
\end{align}
\end{itemize}

\section*{Acknowledgments}
I would like to thank Mario Giovanni Cella for allowing me to closely appreciate the behavioral dynamics of the bee colonies and, together with Alfonso Amendola, for the interesting and useful discussions on the topic covered by this paper. A special thank to Professor Martin Bohner for giving me precious suggestions addressed to improve the quality of the paper.

\bibliographystyle{plain}

\label{lastpage}
\end{document}